# Beyond *The Big Bang Theory*: Revealing the everyday research lives of theoretical physics faculty

Mike Verostek
*Department of Physics and Astronomy, University of Rochester, Rochester, New York 14627, USA*
*School of Physics and Astronomy, Rochester Institute of Technology, Rochester, New York 14623, USA*
Molly Griston
*Department of Physics and Astronomy, University of Rochester, Rochester, New York 14627, USA*
Jesús Botello
*Department of Physics, The University of Texas at Austin, Austin, Texas 78712, USA*
Benjamin Zwickl
*School of Physics and Astronomy, Rochester Institute of Technology, Rochester, New York 14623, USA*

# Introduction

"So what do you do? I mean, what do you actually do?" This is one of the most common questions one theoretical physicist receives from her students, and is representative of a common curiosity among aspiring physicists. Students often misunderstand what different types of physicists actually *do*.[1] For many, the term "theoretical physicist" likely conjures mental images of Einstein, Feynman, or Sheldon Cooper discerning the underlying operating principles of the universe, armed with only pencil and paper. Such imagery is representative of the "lone genius" stereotype many students ascribe to physicists.[2-4] However, this common perception of physicists is known to discourage students, particularly underrepresented gender and racial groups, from pursuing it as a career.[5-8] Moreover, students of all demographic backgrounds may avoid physics if they feel they do not understand what different kinds of physicists do on a daily basis.[9,10] Therefore, providing students with an accurate depiction of the skills, practices, motivations, and daily routines in different types physics careers is essential for helping them make decisions about their future. In this article, we offer students a glimpse into what a career in *theoretical* physics is like. Through interviews with 11 practicing theoretical physics faculty, we reveal a variety of misconceptions theorists believe students hold about their work. Addressing these misconceptions will provide students with important details to consider as they navigate their career decision process.

Studies show that students perceive physicists as more innately brilliant, more socially awkward, and less collaborative than other scientists.[2,5] These stereotypes are even more acute for theorists, who physics students often regard as the most intelligent and esteemed in the physics community.[11] Given that theorists have historically been the most visible and well-known physicists in the popular science sphere,[12,13] offering a nuanced glimpse into the daily lives of typical theoretical physics faculty may be especially illuminating and impactful for prospective physicists. Although several news outlets have published profiles of contemporary theorists,[14,15] little research has been done to systematically characterize what it is like to work as a theoretical physicist.

Previous research articles have shed light on some of the expert practices that physicists, including theorists, use while solving problems in their research.[16-19] However, these articles have largely focused on the problem-solving strategies that these physicists use rather than some of the other aspects of their day-to-day lives. These other aspects include which skills they believe are important, what motivates them in their work, and what a typical day in the life of a practicing theorist looks like. Answers to these questions are essential for students to understand while making career decisions or preparing for a career. For instance, a student motivated to pursue STEM in order to produce new technologies and make an impact on society may overlook theory since many students perceive it as being too abstract.[11] Yet as our results will show, many theorists collaborate tightly with experimentalists and play an instrumental role in producing tangible results. Hence, failing to paint a clear picture of work life across physics-related careers may discourage students from pursuing areas of physics that align most closely with their skills and goals.

To help provide students with a clearer picture of theoretical physics research, we interviewed 11 current theoretical physics faculty members from five colleges and universities. Of the 11 total participants, 6 verbally described their gender as "male," 2 as "female," and 3 as a "woman." We asked these physicists about a variety of topics including their daily activities, skills they use while doing research, and the field of theoretical physics as a whole. Their responses helped us answer two research questions: 1) What do theorists believe are common misconceptions that people, including physics majors, hold about them and their work? and 2) In contrast to these misconceptions, what are the realities of being a theorist and doing theoretical physics? After transcribing the interviews, we analyzed them for common themes related to misconceptions about theory as well as themes related to activities they undertook and attitudes they shared. The results of our analysis are presented here.

## Results

Theorists we interviewed believe that students, the public, and even some other physics faculty often do not have a firm understanding of a theoretical physicist's daily life and responsibilities. As one theorist commented, "I think it's clear to me that [theory] is just not understood. Even sometimes our colleagues, they don't know what we do." Several theorists observed that students occasionally do not differentiate between experiment and theory at all. One recalled, "As an undergraduate, I wasn't really aware of this sort of fork in the road that you choose in graduate school of theory versus experiment. I think most undergraduates… are not so aware of this difference." Yet even students who acknowledge differences between theory and experiment often hold inaccurate ideas about what theorists' work is like, with one theorist positing that this confusion is due in part to the more "abstract" nature of theory. Whereas she believes experimental physics evokes mental images of "doing things with your hands" and "plugging and unplugging things," theoretical physics is harder to picture and can leave students with false impressions of what it is like to be a theoretical physicist.

One of the primary misconceptions that theorists addressed was the student belief that, as one faculty member put it, **theory is "more glamorous" than it really is**. Many interviewees indicated that students think theorists are "just people who talk in fancy terms about deep concepts" without fully appreciating the struggle and long hours associated with theoretical physics research. This misunderstanding is even present among graduate-level physicists, with one faculty member expressing that "Even some people who have worked with me, who are actually very smart and very good at it, they have a false impression that every day you're going to make a breakthrough. The expectations are not realistic." Another offered that "Some students think they like theory because they have seen some science fiction movies. And they say, 'oh, I'm going to go to grad school and I go to be a cosmologist,' and they don't really know what is involved in the work." However, the idealization of theorists as deeply contemplative individuals who routinely produce breakthroughs is a major misrepresentation of their daily work.

In fact, all theorists agreed that the foremost requirement to do theoretical physics research is "blocks of uninterrupted time." Indeed, one theorist tries to give himself three full hours in the afternoon to be "engrossed in research" so that he can read papers thoroughly and do mathematical calculations more carefully. However, opportunities to dedicate many hours to research are easily interrupted by other features of faculty life. All theoretical physics faculty we interviewed characterized their daily lives as being highly variable depending on the day, and broadly split their workdays into a mix of research, teaching, and numerous administrative responsibilities. Tasks they described ranged from serving on committees to filling out hiring paperwork for a new postdoc to "deciding exactly what sort of hors d'oeuvres we should serve at a banquet." On a given day, any of those three activities can take precedence and drown out the others: "There is no typical day in my career."

Despite numerous other responsibilities, theorists recognize that progress in their research is essential for their careers, causing some to seek blocks of time outside the typical workday. And since theorists do not have equipment and can easily work from home, the work-life boundary can become blurred. "Sometimes you have like half an hour or 40 minute blocks between various meetings, but you can't really think on demand like that. So most of the research that I do myself, I have to do it once I come home from work or during weekends. Hence no work-life balance." One theorist advised that trying to maintain that balance while making progress on research is "very demanding on you as a person," while another stressed that "I think for theorists, and also for most researchers, one important thing is to endure uncertainty. Because doing research takes a lot of effort and you have to be patient." Thus, **progress in theory does not always come easily, and is the product of long hours of work dedicated to research**.

This reality relates to another misconception that *all* theorists interviewed sought to dismiss. They indicated that a common but false assumption is that **theoretical physicists are innately brilliant and smarter than other physicists**. Speaking from experience, one theorist recalled "There was this thing when I was an undergrad and we thought theoretical physicists are smarter than everybody else, and experimentalists are not smart. And that is wrong." Another insisted that people "seem to think that [theory] is harder than it actually is," and that "without even trying it, most people somehow dismiss it as too hard for them." One theorist mused, "I'm a person maybe of average intelligence. I'm not consumed by the mysteries of the universe. I mean, they're great. I think they're awesome. The universe is mysterious and incomprehensible, and I'm good with it that way." These quotes indicate that more students are able to contribute to theoretical physics research than they might initially believe.

Indeed, several theorists revealed that since most of their time doing research is spent grappling with a difficult problem, they often do not feel very smart. As one noted, "If you were to ask me on a typical day, I don't think I am particularly good at what I do, because most days you are stuck and you're struggling with something." Hence, contrary to the popular stereotype, nearly all interviewees claimed that **perseverance and endurance were more important character traits of successful theorists than innate brilliance**. Some are willing to work on the same problem for years without giving it up: "I'm not especially good at any of the things that I do, but I am determined and I have tenacity. I go back to it and some of the things that I work on, I work on for years at a time. And that's really the difference. It isn't that I have a special skill or anything else." Another recalled that he spent two "frustrating" years trying an approach to a problem that ended up failing. He said, "My entire professional life from 1995 until 1998 is a footnote in chapter five of my doctoral dissertation that says we tried a different approach and it didn't work," but that he was eventually able to solve the problem through continued effort. Thus, interviewees encouraged students to discount the notion that they have to be a genius to do theory, and to instead try to leverage their strengths to contribute to the field.

Faculty described theory as **a diverse area of physics requiring many different skills and talents**, rather than mere genius. "It requires so many different skills and so many different types of personalities that you always have something to contribute. Don't ever think that you can't do it." Many theorists reported that broad backgrounds in physics and mathematics are helpful in research, but emphasized that being fluent in every single aspect of these disciplines is not a requirement. Reflecting on his own journey to becoming a theorist, one interviewee said that students should "Be aware of stereotypes. And by that I mean, your impression of what it takes to be a good theoretical physicist. So for example, I'm not very good at geometry. Does that disqualify me from being a good theoretical physicist? I don't think so." He continued, "Somebody else who might not even like pen and paper work like I do, but is very good with computers, would make a substantial contribution in a numerical computing project."

Among the diverse skills that theorists described as being important to their success is their ability to collaborate and communicate effectively, which contradicts the image of theorists as **working alone in their office all day.** Several theorists noted that they are commonly thought of as "super smart people who are really good at math and really bad at social skills" and possess little "common sense." Some posited that this image is amplified by popular culture, with one expressing displeasure that people think theorists "are like what they see in Big Bang Theory, and that is wrong. That's a very caricaturized TV show, and I do not like it. I saw a few episodes, and I was like, this is not who I am, this is not who any of my friends are." Indeed, several interviewees said that they have noticed an uptick in people associating them with Sheldon Cooper from The Big Bang Theory.

Contemporary theorists cite their **ability to effectively work with others** as crucial to their research success and their personal happiness on the job.  This reality was summarized by one theorist who said, "What makes me good at what I do - so the thing is, all of my work is collaboration. Even my students who are working with me, I consider them to be my collaborators.  I try to make sure that they have a nice environment to work in, and they're talking to each other and are having a good time. But I think that's my skill."  This interviewee also stated that mentoring is a "big part" of her life and motivates her to push her research forward.

Moreover, many theorists noted that their collaborations extend across disciplines.  One explained that "I try to talk about what I'm working on with as many people as possible. Even if they're not exactly in the same field as me. It's especially useful if they are not in the same field."  This cross-disciplinary communication extends to applied fields such as engineering as well.  Another theorist believes that "One of the things that makes me good at what I do, so the work I do is at the interface of biology and physics and math and engineering. So being able to talk to people who are working in different disciplines and having that comfort and also talking to both theorists and experimentalists, understanding their language, translating my language, which is physics or mathematical physics, to their language, all of that is really, really important."  This reflection alludes to the final misconception that theorists perceived about their work regarding its applicability to real-world problems.

Although many theoretical physicists work on problems that are highly abstract and are unlikely to receive experimental verification any time soon, a subset of theorists we interviewed stressed that **not all theory is "pie in the sky" research**.  Rather, several described their research projects as having direct real-world applications.  As pointed out by one theorist who works on many applied systems, "Many of my friends who are engineers by training but have gone on to do other things tend to think that theoretical physicists nowadays are basically doing pie in the sky research, working on things that could never be seen in experiments realized as machines or devices. So there's some misconception there."

In fact, many theorists **work on projects with practical applications**, and are motivated by the prospect of contributing to the creation of useful devices and applied models.  Five out of the 11 theorists in our sample fell into this category.  Among this group, one noted that she had changed her area of specialization entirely so that she could work on more applied projects. She recalled, "I started out as a string theorist, because I thought I wanted to do string theory. And then within my first year, I was like, this really is too abstract, it's not tangible, and I don't like it."  Indeed, theorists may be deeply integrated into teams of experimentalists working toward a common goal.  Another theorist emphasized, "I really love working with experimentalists. I have a sneaking fondness for them having been one of them when I was a graduate student."  He noted that his willingness to work with experimentalists is not universal among theorists, saying that "A lot of theorists don't like to do that."  Some faculty even considered having the capacity to communicate across disciplinary boundaries as a desirable skill when considering prospective job candidates: "I find when I'm judging a student, or a postdoc, a young faculty member who's a theorist, I look for balance in their work. I find it is important for them to have published theoretical work independent of experiment, but also to have made some predictions which have been verified. My way of thinking, physics is an experimental science, and you can write all the formulas you want, but if they are not verified by some experiment… well, as people say, it's just a theory."

## Discussion and Conclusion

The observations provided by these theorists offer several important insights for both students and educators.

The overarching theme of these interviews for students is that they should not be dissuaded from theory simply because they believe it is reserved for the "smartest" physicists, or that theorists exclusively work in isolation on projects with little relevance to the real world.  Students motivated to do physics due to its collaborative nature or its propensity for producing practical solutions to complex problems may be able to find fulfilling theoretical research opportunities.  In this regard, interviewees portrayed theory as a more broadly accessible research area than many students might initially perceive.  Still, faculty cautioned against unbridled optimism regarding the prospect of becoming a theorist.  Several pointed to the blurred work-life boundary and infrequent breakthroughs in research as aspects of theory that students often find difficult to manage, and students must carefully consider whether such a work structure fits their personality.  Others voiced concerns that there are simply a limited number of research opportunities available in certain areas of theory, particularly popular fields like string theory and quantum gravity, due to funding concerns.  Hence, one faculty member advised prospective students to carefully consider their future goals: "If the interest is predicated on being Sheldon Cooper, don't do it. That's a fantasy. There are worse fantasies one can have for sure, but you're making a career decision, not just a 'what do I want to do?' type decision. You've got to kind of think about what are the tools you're getting for working in a world that's going to change a lot by the time you're in mid-career."  This quote emphasizes the importance of communicating to students a clear picture of what it means to work in different areas of physics, and allowing them opportunities to explore those options for themselves.

In this regard, we hope the observations in this paper prompt educators to reflect on changes they can make in their own classrooms and departments to best support students' career decision-making.  Although much of the high school and undergraduate physics curricula are already geared toward helping students develop their broad math and physics skills, theorists mentioned several skills that may be underemphasized in many classrooms.  Examples include training in interdisciplinary communication, linking theoretical analyses to applied problems, and leveraging computational problem-solving techniques.  Moreover, the reflection offered by most interviewees that their tenacity and perseverance have been the qualities most critical to their success invites consideration of novel means of student assessment.  Further research must be done to reckon with how to best incorporate features of authentic research practices into classroom instruction, as well as assessing whether our procedures of evaluating students emphasize a sufficiently wide breadth of student skills and attitudes.  Such work will be essential for sustaining a diverse and multitalented physics workforce, and we encourage physics educators to use the topics presented in this paper as inspiration for better supporting student career decision-making.

# References

1. R. Bennett, D.Z. Alaee, and B. Zwickl, Analysis of physics students' subfield career decision-making using social cognitive career theory, in *Physics Education Research Conference 2022*, PER Conference (Grand Rapids, MIC, 2022) pp. 51-56.
2. M. Bruun, S. Willoughby, and J. L. Smith, Identifying the stereotypical who, what, and why of physics and biology, Physical Review Physics Education Research **14**, 020125 (2018).
3. B. Wong, Y. T. Chiu, Ó. M. Murray, J. Horsburgh, and M. Copsey-Blake, 'Biology is easy, physics is hard': Student perceptions of the ideal and the typical student across STEM higher education, International Studies in Sociology of Education, 1 (2022).
4. M. A. Weitekamp, The image of scientists in The Big Bang Theory, Physics Today **70**, 40 (2017).
5. S. Leslie, A. Cimpian, M. Meyer, and E. Freeland, Expectations of brilliance underlie gender distributions across academic disciplines, Science **347**, 262 (2015).
6. A. J. Gonsalves, "Physics and the girly girl - There is a contradiction somewhere": Doctoral students' positioning around discourses of gender and competence in physics, Cultural Studies of Science Education **9**, 503 (2014).
7. Z. Y. Kalendar, E. Marshman, C. D. Schunn, T. J. Nokes-Malach, C. Singh, Framework for unpacking students' mindsets in physics by gender, Physical Review Physics Education Research **18**, 010116 (2022)
8. A. Johnson, M. Ong, L. T. Ko, J. Smith, and A. Hodari, Common challenges faced by women of color in physics, and actions faculty can take to minimize those challenges, The Physics Teacher **55**, 356 (2017).

9. D. Farland-Smith, How does culture shape students' perceptions of scientists? Cross-national comparative study of American and Chinese elementary students, Journal of Elementary Science Education, **21**, 23, (2009).

10. E. Parisi, G. Masia, C. Reynolds, and A. Richards, Investigating High School Students' Perception of Careers and Diversity Issues Within Physics, The Physics Teacher **61**, 140 (2023).

11. A. Johansson, Negotiating intelligence, nerdiness, and status in physics master's studies, Research in Science Education **50**, 2419 (2020).

12. A. Jogalekar, Theorists, experimentalists, and the bias in popular physics, Available at https://blogs.scientificamerican.com/the-curious-wavefunction/popular-physics-is-there-an-experimentalist-in-the-house/ (2013).

13. C. Orzel, American Physicists and the Under-rating of Experiments, Available at https://scienceblogs.com/principles/2013/05/21/american-physicists-and-the-under-rating-of-experiments (2013).

14. N. Collins, What it's like to be a theoretical physicist, Available at https://news.stanford.edu/2019/05/03/whats-like-theoretical-physicist/ (2019).

15. H. Jarlett, Why bother with theoretical physics?, Available at https://home.cern/news/series/in-theory/theory-why-bother-theoretical-physics (2016).

16. M. Verostek, M. Griston, J. Botello, and B. Zwickl, Making expert processes visible: How and why theorists use assumptions and analogies in their research, Physical Review Physics Education Research **18**, 020143 (2022).

17. M. Verostek, M. Griston, J. Botello, and B. Zwickl, Making expert cognitive processes visible: planning and preliminary analysis in theoretical physics research, in *Physics Education Research Conference 2022*, PER Conference (Grand Rapids, MI, 2022) pp. 469-474.

18. A. M. Price, C. J. Kim, E. W. Burkholder, A. V. Fritz, and C. E. Wieman, A detailed characterization of the expert problem-solving process in science and engineering: Guidance for teaching and assessment, CBE-Life Sciences Education **20**, ar43 (2021).

19. M. Griston, J. Botello, M. Verostek, and B. Zwickl, When the light bulb turns on: motivation and collaboration spark the creation of ideas for theoretical physicists, in *Physics Education Research Conference 2021*, PER Conference (Virtual Conference, 2021) pp. 160-156.